\documentclass[10pt]{article}
\usepackage{amsmath,amssymb,latexsym}
\usepackage{graphicx}

\advance \topmargin by -\headheight
\advance \topmargin by -\headsep
     
\evensidemargin \oddsidemargin
\newcommand\rf[1]{(\ref{eq:#1})}
\newcommand\lab[1]{\label{eq:#1}}
\newcommand\nonu{\nonumber}
\newcommand\br{\begin{eqnarray}}
\newcommand\er{\end{eqnarray}}
\newcommand\be{\begin{equation}}
\newcommand\ee{\end{equation}}

\newcommand\foot[1]{\footnotemark\footnotetext{#1}}
\newcommand\lb{\lbrack}
\newcommand\rb{\rbrack}
\newcommand\llangle{\left\langle}
\newcommand\rrangle{\right\rangle}

\newcommand\llb{\left\lbrack}
\newcommand\rrb{\right\rbrack}

\newcommand\lcurl{\left\{}
\newcommand\rcurl{\right\}}
\renewcommand\({\left(}
\renewcommand\){\right)}
\newcommand\bv{\bigm\vert}               
\newcommand\Bgv{\;\Bigg\vert}            

\newcommand\bc{\begin{center}}
\newcommand\ec{\end{center}}

\newcommand\Tr{\mathop{\mathrm Tr}}                  
\newcommand\partder[2]{\frac{{\partial {#1}}}{{\partial {#2}}}}

\renewcommand\a{\alpha}

\renewcommand\d{\delta}
\newcommand\D{\Delta}

\newcommand\vareps{\varepsilon}
\newcommand\g{\gamma}
\newcommand\G{\Gamma}

\newcommand\h{\frac{1}{2}}
\renewcommand\k{\kappa}
\renewcommand\l{\lambda}

\newcommand\m{\mu}
\newcommand\n{\nu}

\newcommand\vp{\varphi}
\renewcommand\P{\Phi}
\newcommand\pa{\partial}

\renewcommand\r{\rho}
\newcommand\s{\sigma}
\renewcommand\S{\Sigma}
\renewcommand\t{\tau}

\newcommand\ti{\tilde}

\newcommand\cD{{\mathcal D}}

\newcommand\cJ{{\mathcal J}}

\newcommand\cL{{\mathcal L}}

\newcommand\cN{{\mathcal N}}

\newcommand\cU{{\mathcal U}}

\begin{document}

\title{Large $N$ Expansion. Vector Models
\foot{Preliminary version of a contribution to the ``Quantum Field
Theory. Non-Perturbative QFT'' topical area of 
{\em Modern Encyclopedia of Mathematical Physics (SELECTA)}, 
eds. Aref'eva I, and Sternheimer D, Springer (2007)}
}
\author{Emil Nissimov and Svetlana Pacheva\\
\small\it Institute for Nuclear Research and Nuclear Energy,\\[-1.mm]
\small\it Bulgarian Academy of Sciences, Sofia, Bulgaria  \\[-1.mm]
\small\it email: nissimov@inrne.bas.bg, svetlana@inrne.bas.bg}
\date{ }
\maketitle

\underline{Large-$N$ expansion} of quantum field theory (QFT) models with 
internal $U(N)$ (or $O(N)$) ``flavor'' symmetry,  where the fundamental matter 
fields belong to the vector representation ({\em vector QFT models} for short), 
is one of the principal, most well 
understood and systematically developed non-perturbative methods. Here the
term ``non-perturbative'' means that \underline{large-$N$ expansion} is qualitatively 
different from standard (``naive'') \underline{perturbation theory} w.r.t. coupling 
constant(s), as its diagrams involve new types of internal propagator lines
corresponding to auxiliary ``flavor''-singlet composite fields which are given
by infinite resummation of (subsets of) ordinary 
\underline{Feynman diagrams}. The latter 
results in improved ultraviolet (UV) behavior of large-$N$ diagrams which 
coupled with the fact that $1/N$ is dimensionless expansion parameter makes 
\underline{renormalizable} those vector QFT models which 
are non-renormalizable w.r.t. the ordinary \underline{perturbation theory}.
Another important general property of \underline{large-$N$ expansion} is that it
exhibits {\em linear} realization of $U(N)$ (or $O(N)$) ``flavor'' symmetry 
in QFT models with a {\em nonlinearly} realized ``flavor'' symmetry,
\textsl{i.e.}, the \underline{nonlinear sigma-models}.

Large-$N$ expansion is the main instrument in uncovering and for explicit
description of the following important properties of QFT
(henceforth dimensionality of space-time will be denoted by $D$):

(i) $D=2$ QFT: dynamical breakdown of classical \underline{conformal symmetry}
via \underline{dimensional transmutation} of coupling constants together with 
\underline{asymptotic freedom}, as well as construction of 
\underline{higher local quantum conserved} \underline{currents} in $D=2$ 
\underline{integrable models}.

(ii) $D\geq 3$ QFT:
dynamical breaking of continuous and discrete (space- and time-reflection)
symmetries;
non-trivial phase structure (several distinct types of phases with multiple
\underline{order parameters}) and the pertinent \underline{phase transitions};
non-perturbative particle spectra qualitatively different in the various
phases;
\underline{dynamical mass generation} for the fundamental $N$-component matter 
fields; dynamical generation of massive gauge bosons where the standard 
\underline{Higgs mechanism}
is inoperative; particle confinement in some of the phases, explicit appearance
of composite bosons and fermions; renormalization of non-renormalizable (w.r.t.
ordinary \underline{perturbation theory}) QFT models.

(iii) Further applications of \underline{large-$N$ expansion} of vector models in various
areas of QFT and statistical mechanics (\textsl{i.e.}, \underline{Euclidean QFT})
include: 
finite-size effects (finite-size scaling in the nonlinear sigma-models); 
\underline{stochastic Langevin equation} in dissipative dynamics;
\underline{finite-temperature QFT} (dimensional reduction crossover at 
high temperature);
\underline{Bose-Einstein condensation} in weakly interacting Bose gas;
\underline{multicritical} \underline{points} and \underline{double scaling limit};
for a comprehensive review, see ref.[1].

Derivation of \underline{large-$N$ expansion} via \underline{functional integral}
techniques is based on the following general prescription:
(a) Introduce appropriate set of auxiliary ``flavor''-singlet fields and
rewrite the original action in a (classically) equivalent form which is
quadratic w.r.t. fundamental $N$-component matter fields; 
(b) In the \underline{functional integral} expression for the 
\underline{generating functional} of the \underline{quantum correlation functions}
perform the \underline{Gaussian functional integral} 
over the $N$-component matter fields to obtain
an effective action depending only on ``flavor''-singlet fields, where the 
factor $N$ appears as a common factor in front of it in the same way as the
\underline{Planck constant} appears as a common factor $1/\hbar$ in front of
the ordinary classical action in the standard 
\underline{functional integral};
(c) Then the \underline{large-$N$ expansion} becomes \underline{``semiclassical'' expansion}
around saddle points of the effective ``flavor''-singlet action, which can be
viewed as \underline{vacuum expectation values} of the pertinent 
``flavor''-singlet fields in the leading order w.r.t. $1/N$ 
(cf. Eqs.\rf{vev} below).

Our first example will be the \underline{large-$N$ expansion} in $D=2$ $O(N)$
\underline{nonlinear sigma-model} whose Lagrangian is given by:
\vspace{-0.3cm}
\be
\cL_{NLSM} = \pa_{+} \vp^a \pa_{-} \vp_a , \qquad {\vec{\vp}}^{\; 2} = N/g 
\quad ,\quad \pa_{\pm} = \h \(\partder{}{x^0} \pm \partder{}{x^1}\)
\qquad \vec{\vp} = \( \vp^1 ,\ldots \vp^N \)  
\lab{NLSM}
\ee
The \underline{large-$N$ expansion} is obtained from the \underline{generating functional}
of the time-ordered correlation functions: 
\br
Z \lb J\rb = \int \cD\vec{\vp}\, \prod_x \d \( {\vec{\vp}}^{\; 2} - N/g \) \,
\exp \lcurl i\int d^2 x \, \llb \pa_{+}\vec{\vp}\, \pa_{-}\vec{\vp}
+ \(\vec{J},\vec{\vp}\) \rrb \, \rcurl   
\lab{n-action} \\
= \int \cD\vec{\vp}\,\cD{\a}\,\exp \lcurl i\int d^2 x \, 
\llb \pa_{+}\vec{\vp}\, \pa_{-}\vec{\vp} - \h \a \( {\vec{\vp}}^{\; 2} - N/g \)
+ \(\vec{J},\vec{\vp}\) \rrb \, \rcurl
\nonu  \\
= \int \cD \a \,\exp \lcurl - \frac{N}{2} S_1 \lb \a\rb +
\frac{i}{2} \int d^2 x\, d^2 y\, \( \vec{J}(x), (-\pa^2 + \a )^{-1} \vec{J}(y)\)
\,\rcurl 
\nonu  \\
S_1 \lb \a\rb \equiv \Tr \ln (-\pa^2 + \a ) - \frac{i}{g} \int d^2 x \, \a
\quad ,\quad \pa^2 = - \(\partder{}{x^0}\)^2 + \(\partder{}{x^1}\)^2
\lab{eff-action}
\er
by expanding the effective $\a$-field action \rf{eff-action} around its constant
saddle point ${\widehat \a} \equiv m^2$, \textsl{i.e.}, 
$\a (x) = m^2 + \frac{1}{\sqrt{N}} {\ti \a}(x)$. From the stationary equation 
$\delta S_1 \lb \a\rb / \delta \a \bv_{\a = m^2} = 0$ one obtains 
$m^2 = {\hat\m}^2 e^{-4\pi/g}$, where ${\hat\m}$ is a renormalization mass scale
appearing due to \underline{renormalization} of the UV divergence coming from 
the first term in \rf{eff-action} (see Eq.\rf{ren} below). 
Thus, the \underline{``Goldstone'' fields} $\vec{\vp}$
acquire dynamically generated mass (squared) ${\widehat \a} \equiv m^2$, 
classical \underline{conformal invariance} of \rf{n-action} is broken due to the
\underline{dimensional transmutation} (the dimensionless coupling $g$ is 
replaced by $m^2$), and the classically nonlinearly realized $O(N)$ ``flavor''
symmetry becomes linearly realized on the quantum level.
From \rf{eff-action} one arrives at the large-$N$ \underline{diagram technique}
with (free) propagators in momentum space:
$\llangle \vp^a \, \vp^b \rrangle_{(0)} = -i \( m^2 + p^2 \)^{-1} \d^{ab}$, 
$~\llangle {\ti \a} \, {\ti \a} \rrangle_{(0)} = \Bigl( \S \(p^2 \) \Bigr)^{-1}$
with
$\S \(p^2\) = \int\frac{d^2 k}{(2\pi)^2}\,
\llb\( m^2 +k^2\)\( m^2 +(p-k)^2\)\rrb^{-1}$,
and tri-linear ${\ti\a}\vp\vp$-vertices, where one-loop $\vp$-tadpoles and 
subdiagrams of the form in the picture below are forbidden (solid lines depict
$\vp$ propagators, dashed lines depict ${\ti\a}$ propagators). 
\begin{center}
\includegraphics[scale=0.5]{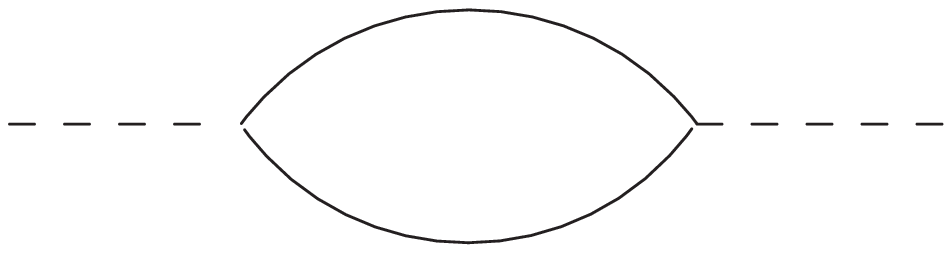}
\end{center}

The diagrams of the \underline{large-$N$ expansion} of vector QFT models
still contain UV divergences which can be
systematically renormalized 
both in $D=2$ and $D\geq 3$ by a
version of the mass-independent (``soft-mass'') momentum-space subtraction 
procedure of Zimmermann-Lowenstein, which in turn is based on 
\underline{BPHZ} (Bogoliubov-Parasiuk-Hepp-Zimmermann) renormalization scheme. 
The mass-independent momentum-space subtraction renormalization (for details,
see the link \underline{BPHZL Renormalization}) has the advantage over other
renormalization schemes that it 
can be applied simultaneously and in an uniform way in all phases of the 
pertinent QFT models, especially in those of them with 
phases containing massless particles where particular care is needed to avoid
possible \underline{infra-red singularities}.

A remarkable property of the \underline{large-$N$ expansion} in nonlinear sigma models is
that the nonlinearity of the \underline{``Goldstone'' field}
$\vec{\vp}(x)$ is preserved on
the quantum level as an identity on the correlation functions, in spite
of the manifest linear $O(N)$ symmetry of the large-$N$ diagrams:
\be
\llangle \cN \Bigl\lb {\vec{\vp}}^2 P(\vec{\vp}, \pa \vec{\vp}) \Bigr\rb (x)
\ldots \rrangle =
{\text{const}} \llangle \cN \Bigl\lb P(\vec{\vp}, \pa \vec{\vp}) \Bigr\rb (x)
\ldots \rrangle                   
\lab{chiral}
\ee
where $P(\vec{\vp}, \pa \vec{\vp})\,$ is arbitrary local polynomial 
of the fundamental fields and their derivatives, and $\cN \lb \ldots \rb\,$ 
indicates \underline{BPHZL}-renormalized \underline{normal product}
of the corresponding composite fields.

Using the \textsl{BPHZL}-renormalized \underline{large-$N$ expansion} one can explicitly
construct the higher quantum local conserved currents $\cJ^{(s)}_{\pm}$
for the model \rf{NLSM} 
($\pa_{+}\cJ^{(s)}_{-} + \pa_{-}\cJ^{(s)}_{+} = 0$, $s=3,5,\ldots$, where
$s$ indicates the $D=2$ Lorentz spin of the corresponding higher conserved
charge). Their existence is of profound importance as they imply 
\underline{quantum integrability} of the $O(N)$ nonlinear sigma-model \rf{NLSM}.
The first non-trivial \underline{higher quantum local conserved current}
is of the form:
\be
\cJ^{(3)}_{-} = \cN \llb\( \pa_{-}^2 \vec{\vp} \)^2 \rrb +
a_1 \cN \llb\( (\pa_{-} \vec{\vp} )^2 \)^2 \rrb \quad ,\quad 
\cJ^{(3)}_{+} = \( \h + a_2 \) \cN \llb\(\pa_{-} \vec{\vp} \)^2 \a\rrb +
a_3 \pa_{-}^2 \a  
\lab{3-current}
\ee
where all coefficients $a_{1,2,3} = O (1/N)\,$ are expressed in terms of 
one-particle irreducible correlation functions and their derivatives in 
momentum space at zero external momenta. Their explicit form can be found order
by order in $1/N$ from the renormalized large-$N$ diagram technique described
above. Let us stress, that \rf{3-current} as well as its higher counterparts
$\cJ^{(s)}_{\pm}$ for $s=5,7,\ldots\,$ do not have analogues in the classical 
conformally invariant $O(N)$ nonlinear sigma-model \rf{NLSM}.

As a second non-trivial example let us consider the $D=3$ gauged 
\underline{nonlinear sigma-models} with fermions $(GNLSM+F)_3$, with internal
symmetry $U(N)(\textrm{``flavor''}) \times U(n)(\textrm{``color'' gauge})$ 
($n<N$).
Special physically relevant cases of the latter are the 
\underline{supersymmetric} \underline{nonlinear sigma-models}
on complex projective and Grassmannian manifolds
(\textsl{e.g.}, by taking $e^2 \to \infty$, $\vareps =1$ in \rf{GNLSM+F} below).
On the other hand, $(GNLSM+F)_3$ themselves can be viewed as special fixed
points of general $D=3$ (non-Abelian) Higgs models with fermions, 
containing ``non-renormalizable'' four-fermion couplings.
$(GNLSM+F)_3$ are of particular physical interest since their large-$N$
expansion explicitly displays all the fundamental non-perturbative
properties listed above under (ii). Below we will discuss for simplicity 
\underline{large-$N$ expansion} for the Abelian $(GNLSM+F)_3$ ($n=1$, the non-Abelian case
being a straightforward generalization).

The pertinent Lagrangian reads:
\be
\cL_{\textrm{GNLSM+F}} = - \(\nabla^{\n}(A) \vp_a\)^{*} \(\nabla_{\n}(A) \vp^a\)
+ i{\bar\psi}_a \g^\n\nabla^{(\vareps)}_{\n}(A)\psi^a + 
\frac{g}{4N\m}\({\bar\psi}_a \psi^a\)^2
- \frac{N}{4e^2\m} F_{\n\l}(A) F^{\n\l}(A)
\lab{GNLSM+F}
\ee
with constraints 
$\vp^{*}_a\vp^a - N\m/T =0 \quad,\quad {\bar\psi}_a\vp^a = \vp^{*}_a\psi^a =0$.
Here the following notations are used:
$\nabla_{\n}(A) \vp^a = \(\pa_\n + i A_\n\)\vp^a$, 
~$\nabla^{(\vareps)}_{\n}(A) \psi^a = \(\pa_\n + i \vareps A_\n\)\psi^a$,
~$F_{\n\l}(A) = \pa_\n A_\l - \pa_\l A_\n$, where the ``flavor'' indices 
$a=1,\ldots,N$ and the space-time indices $\n,\l = 0,1,2$; 
$\vareps$ is the ratio of fermionic to bosonic electric charges;
$\g_\n$ are the standard $D=3$ Dirac gamma-matrices;
$\m$ denotes a common mass scale exhibiting the dimensionfull nature of the
coupling constants $T/\m, g/\m, e^2\m$. Note the presence of the
``non-renormalizable'' four-fermion (\underline{{\em Gross-Neveu}}) term 
in \rf{GNLSM+F}.

Apart from the continuous 
$U(N)(\textrm{``flavor''}) \times U(1)(\textrm{gauge})$ symmetry, 
$(GNLSM+F)_3$ \rf{GNLSM+F} is invariant also under the discrete space-time
transformations -- space ($P$-) and time ($T$-) reflections:
$\vp^{(P,T)} (x) = \eta_{P,T} \vp (x_{P,T})$, 
$\psi^{(P,T)} (x) = \eta_{P,T} \g_{1,2}\vp (x_{P,T})$,
$A^{(P)} (x) = (A_0,-A_1,A_2)(x_{P,T})$,
$A^{(T)} (x) = (A_0,-A_1,-A_2)(x_{P,T})$, where
$x_P = (x^0,-x^1,x^2)$, $x_T = (-x^0,x^1,x^2)$ and $|\eta_{P,T}|=1$.
Note that fermionic mass term reverses sign under $P,T$-reflection:
$\bar{\psi}^{(P,T)} \psi^{(P,T)} (x)= - \bar{\psi}\psi (x_{P,T})$ and due to
its absence in \rf{GNLSM+F} the classical $(GNLSM+F)_3$ is $P,T$-invariant.
Therefore, $P,T$-reflection symmetries can be viewed as $D=3$ analogues of the 
\underline{chiral symmetry} in $D=4$ gauge theories with massless chiral fermions. 

Introducing a set of auxiliary $U(N)$-singlet fields (real scalar $\a$,$\s$
and complex fermionic $\rho$) one can rewrite the action \rf{GNLSM+F} in
the following (classically) equivalent form:
\br
L_{\textrm{GNLSM+F}} = - \(\nabla^{\n}(A) \vp_a\)^{*} \(\nabla_{\n}(A) \vp^a\)
- \a \(\vp^{*}_a\vp^a - N\m/T\) + 
i{\bar\psi}_a \g^\n\nabla^{(\vareps)}_{\n}(A)\psi^a - \s {\bar\psi}_a \psi^a 
\nonu \\
- \frac{N\m}{g}\,\s^2 + \vp^a \({\bar\psi}_a\rho\) + 
\({\bar\rho}\psi^a\)\vp^{*}_a - \frac{N}{4e^2\m} F_{\n\l}(A) F^{\n\l}(A)
\lab{GNLSM+F-1}
\er

Derivation of the \underline{large-$N$ expansion} for the quantum 
\underline{generating functional} $Z[J_\P]$ 
of \rf{GNLSM+F-1} proceeds along similar lines as for the $D=2$ $O(N)$
\underline{nonlinear sigma-model} 
\rf{NLSM}--\rf{eff-action}. Unlike the $D=2$ case, in $D\geq 3$ the
fundamental $N$-component scalar field may acquire non-zero \underline{vacuum
expectation value} for certain range of the parameters, therefore, it is
appropriate to split it in two parts -- parallel and orthogonal w.r.t.
direction of the (possible) \underline{vacuum expectation value}: 
$\vec{\vp} = \vec{\vp}_{||} + \vec{\vp}_{\perp}$.
Without loss of generality one may choose 
$\vec{\vp}_{||} = (0,\ldots,0,N^{\h}\vp_{||})$
and $\vec{\vp}_{\perp} = (\vp_1,\ldots,\vp_{N-1},0)$. Then performing the
\underline{Gaussian functional integration} w.r.t. $\vec{\vp}_{\perp}$, $\psi$ 
one gets:
\br
Z[J_\P] =\int\cD\vec{\vp}_{\perp}\cD\psi\cD\vp_{||}\cD\a\cD\s\cD\rho\cD A_{\m}\,
\exp\Bigl\{ i \int d^3 x \Bigl\lb L_{\textrm{GNLSM+F}} + 
\sum_{\P = \vp,\psi,\ldots} J_\P (x) \P (x)\Bigr\rb\Bigr\}
\lab{GNLSM+F-2a} \\
= \int \cD\vp_{||} \cD\a \cD\s \cD\rho \cD A_{\m}\,
\exp\Bigl\{ iNS_1\llb \vp_{||},\a,\s,\rho,A\rrb + iS_2[J_\P]\Bigr\}
\lab{GNLSM+F-2}
\er
Here the effective action reads: 
\br
S_1\llb \vp_{||},\a,\s,\rho,A\rrb = i(1-1/N) \Tr\ln\D_B - i \Tr\ln\D_F
\nonu \\
+ \int d^3 x \Bigl\lb - \h \vp^{*}_{||} \D_B \vp_{||} - \a\, \m/T
- \s^2 \m/g - \frac{1}{4e^2\m} F_{\n\l}(A) F^{\n\l}(A) \Bigr\rb
\lab{GNLSM+F-3}
\er
where $\D_F \equiv i\g^\l \nabla^{(\vareps)}_\n (A) - \s$, 
~$\D_B \equiv - \nabla^{\n} (A) \nabla_{\n} (A) + \a + \bar{\rho} \D_F^{-1} \rho$, 
and $S_2[J_\P]$ contains the terms with the sources.

Because of \underline{Lorentz invariance} of the vacuum 
only $\vp_{||}$, $\a$ and $\s$ may have non-zero constant stationary values
${\widehat \vp}_{||} \equiv v$, ${\widehat \a} \equiv m_{\vp}^2$, 
${\widehat \s} \equiv m_{\psi}$ where:
\be
\llangle \vp^a \rrangle = N^{\h}\llb v\d^a_N + O(N^{-1})\rrb \;, \;
\llangle \a \rrangle = m_{\vp}^2 + O(N^{-1}) \; ,\;
\llangle {\bar\psi}\psi \rrangle = \frac{2N\m}{g} \llangle \s \rrangle
= \frac{2N\m}{g}\llb m_{\psi} + O(N^{-1})\rrb
\lab{vev}
\ee
Thus, the saddle-point equations acquire the form:
\br
\frac{\d S_1}{\d \vp^{*}_{||}} = - m_{\vp}^2\, v = 0  \; ,\;
\frac{\d S_1}{\d \a} = 
\frac{m_{\vp}}{4\pi} - \llb |v|^2 + \m \(\frac{1}{T_c} - \frac{1}{T}\)\rrb = 0
\; ,\;
\frac{\d S_1}{\d \s} = 
- 2m_{\psi} \llb\frac{m_{\psi}}{4\pi} - \m \(\frac{1}{T_c} - \frac{1}{g}\)\rrb = 0
\lab{saddle-3}
\er
The dimensionless constant $T_c = 4\pi \m/{\hat \m}$ arises in the evaluation 
of the UV-divergent integrals appearing in the variational derivatives of $S_1$ 
which are renormalized according to the ``soft-mass'' \underline{BPHZL}
subtraction scheme with arbitrary scale ${\hat\m}$ (in particular, one may 
take ${\hat\m} = \m$) :
\br
i\frac{\d \Tr\ln\D_B}{\d \a}
\Bgv_{{\widehat \a}=m_{\vp}^2,\ldots,{\widehat \rho}=0} = 
\lcurl i \int d^D p/(2\pi)^D \,\lb m^2_{\vp} + p^2\rb^{-1} \rcurl^{\textrm{ren}} 
\nonu \\
= i \int d^D p/(2\pi)^D \,\llb \( m^2_{\vp} + p^2\)^{-1} -
\( {\hat\m}^2 + p^2\)^{-1}\rrb = \left\{ 
\begin{array}{ll}
\frac{1}{4\pi} \ln\( m_{\vp}^2/{\hat\mu}^2\) \quad & \;\; \textrm{for}\; D=2 \\
\frac{1}{4\pi} \( m_{\vp} - {\hat\mu}\) \quad & \;\; \textrm{for}\; D=3
\end{array} \right.
\lab{ren}
\er
and similarly for 
$-i \lcurl\d \Tr\ln\D_F/\d \s \rcurl\bv_{{\widehat\s}=m_{\psi},A=0}$~.

The solutions of the saddle-point equations \rf{saddle-3} yield the
following phase structure of $(GNLSM+F)_3$ \rf{GNLSM+F} characterized by
{\em two} \underline{order parameters} $\llangle \vec{\vp} \rrangle$,
$\llangle {\bar\psi}\psi \rrangle$ \rf{vev}:

(I) $U(N)(\textrm{``flavor''}) \times U(1)(\textrm{gauge})$ and $P,T$-symmetric
``high-temperature'' phase for $T>T_c$ and $0<g<T_c$, where: $v=0$, 
$m_{\vp} = 4\pi\m \( 1/T_c - 1/T\)$, $m_{\psi} = 0$.

(II) $U(N)(\textrm{``flavor''}) \times U(1)(\textrm{gauge})$ symmetric
``high-temperature'' phase with \underline{spontaneous breakdown} of discrete
$P,T$-reflection symmetries due to dynamical generation of fermionic mass
$m_{\psi}$ for $T>T_c$ and either $g<0$ or $T_c < g < 2T_c$, where:
$v=0$, $m_{\vp} = 4\pi\m \( 1/T_c - 1/T\)$, $m_{\psi} =4\pi\m \( 1/T_c - 1/g\)$.

(III) $P,T$-symmetric ``low-temperature'' phase with \underline{spontaneous 
breakdown}
of internal $U(N)(\textrm{``flavor''}) \times U(1)(\textrm{gauge})$ due to
non-zero $\llangle\vec{\vp}\rrangle$ \rf{vev}
for $T<T_c$ and $0<g<T_c$, where:
$|v|^2 = \m \( 1/T - 1/T_c\)$, $m_{\vp} = 0$, $m_{\psi} = 0$.

(IV) ``Low-temperature'' phase with \underline{spontaneous breakdown}
of both the discrete
$P,T$-symmetries (as in phase (II)) and internal symmetry (as in phase (III))
for $T<T_c$ and either $g<0$ or $T_c < g < 2T_c$, where:
$|v|^2 = \m \( 1/T - 1/T_c\)$, $m_{\vp} = 0$, 
$m_{\psi} =4\pi\m \( 1/T_c - 1/g\)$.

Let us recall that $P,T$-reflection symmetries are $D=3$ analogues of the
fermionic \underline{chiral symmetry} in $D=4$. 

The restriction $g < 2T_c$ above originates from the stability requirement
for the \underline{quantum effective potential} of $(GNLSM+F)_3$ \rf{GNLSM+F}.
According to the general definition it is given as a Legendre transform of
the logarithm of the quantum generating functional \rf{GNLSM+F-2a}:
$\cU \(\langle\vec{\vp}\rangle, \langle\bar{\psi}\psi\rangle\) = 
-i \ln Z\llb J_{\vp},J_{\bar{\psi}\psi}\rrb - 
\( J^{*}_{\vp\, a} \langle\vp^a\rangle + \langle\vp^{*}_a\rangle J_{\vp}^a
+ J_{\bar{\psi}\psi} \langle\bar{\psi}\psi\rangle\)$.
In the large-$N$ limit one obtains (cf. the relations \rf{vev}):
$N^{-1}\cU \(\langle\vec{\vp}\rangle, \langle\bar{\psi}\psi\rangle\) =
\cU_1 \(\langle\vec{\vp}\rangle,\langle\s\rangle\) -
g/4\m\, \(\d \cU_1/\d \langle\s\rangle\)^2$ where
$\cU_1 \(\langle\vec{\vp}\rangle,\langle\s\rangle\) =
1/6\pi\, \(|\langle\s\rangle|^3 - \langle\a\rangle^{3/2}\)
- \m |\langle\s\rangle|^2 \( 1/T_c - 1/g\) + 
\langle\s\rangle \llb |\langle\vec{\vp}\rangle|^2 + \m \( 1/T_c - 1/T\)\rrb$.

All transitions between any pair of the above phases are 
\underline{second-order} on the lines $T=T_c$ and $g=T_c$ on the $(T,g)$ 
parameter plane. On the other hand, the line $g=0$ corresponds to 
\underline{first-order} phase transitions between phases 
(I) and (II) for $T>T_c$, and between phases (III) and (IV) for $T<T_c$.

All four phases exhibit qualitatively different non-perturbative particle
spectra. The spectra are directly derived from the momentum-space pole structure
of the propagators in the pertinent large-$N$ diagrams, where the
propagators themselves are determined from the quadratic part of the
expansion of the large-$N$ effective action \rf{GNLSM+F-3} around its 
saddle points. The highlights of these spectra include appearance of composite
bosons and fermions in phases (II) and (IV), \underline{``confinement''}
of part of the fundamental $N$-component matter fields ($\vp$, $\psi$) in 
phases (III) and (IV).
The most interesting effect occurs in phase (II), which contains massive gauge
bosons due to dynamical generation in the large-$N$ effective action 
\rf{GNLSM+F-3} of a $P,T$-noninvariant \underline{topological Chern-Simmons term} 
$1/16\pi \,\int\! d^3\! x\, \vareps^{\k\l\n} A_{\k} F_{\l\n}(A)$. 
In all other phases the gauge bosons are ``confined'' due to 
appearance of $\sqrt{p^2}$-singularity in the $A_\n$-propagators. Thus, in
spite of the unbroken gauge symmetry in phases (I) and (II), massless gauge
bosons are absent there. Also, the standard \underline{Higgs mechanism}
for generating masses of gauge bosons does not operate in phases (III) and (IV)
in spite of the \underline{spontaneous breakdown} of the gauge symmetry there.

It is also worth mentioning that at the critical point $T=T_c,\, g = T_c$
and upon taking the scaling limit $(GNLSM+F)_3$ \rf{GNLSM+F} becomes the
$D=3$ \underline{supersymmetric non-linear sigma-model} on the complex
projective space $CP^{N-1}$: $
\cL_{\textrm{susy} CP^{N-1}} = 
- \(\nabla^{\n}(A) \vp_a\)^{*} \(\nabla_{\n}(A) \vp^a\)
+ i{\bar\psi}_a \g^\n\nabla_{\n}(A)\psi^a + 
\frac{T_c}{4N\m}\({\bar\psi}_a \psi^a\)^2$ with constraints 
$\vp^{*}_a\vp^a - N\m/T_c =0 \quad,\quad {\bar\psi}_a\vp^a = \vp^{*}_a\psi^a =0$.
This is a {\em non-trivial $D=3$} \underline{conformal field theory} with a 
well-defined renormalizable \underline{large-$N$ expansion} where all relevant 
\underline{anomalous conformal dimensions} (some of them describing the 
\underline{critical behaviour} of $(GNLSM+F)_3$ \rf{GNLSM+F} in the vicinity 
of the \underline{second-order phase transitions}) can be
explicitly computed order by order in $1/N$ from the large-$N$ diagram
techniques.

\vspace{.3cm}
\textbf{Further Reading.}

Ref.[1] contains a comprehensive review (together with an extensive list of 
references) of most of the relevant aspects and applications of large-$N$ 
expansion of vector QFT models, especially those mentioned under (iii) above
(see also the book [2]).
More details about application of \underline{large-$N$ expansion} to 
construct higher local quantum conserved currents in $D=2$ integrable QFT
models with $O(N)$ (or $U(N)$) internal symmetry can be found in refs.[3,4]
and references therein.
For a systematic renormalization of the \underline{large-$N$ expansion}, including
proofs of renormalizability of QFT models which are non-renormalizable within the
standard \underline{perturbation theory}, see refs.[5,6,7] and references therein. 
Further details about application of \underline{large-$N$ expansion} to derive non-trivial
phase structure and non-perturbative particle spectra in $D=3$ gauge theories 
with fermions, including supersymmetric nonlinear sigma-models, can be found in
refs.[7,6] and references therein. For the role of large-$N$ vector QFT models
in the context of anti-de-Sitter/conformal-field-theory dualities in modern
non-perturbative string theory, see ref.[8].

\vspace{.3cm}
[1] Moshe M, and Zinn-Justin J (2003)
{\em Quantum Field Theory in the Large N Limit: a Review}.
Physics Reports 385:69-228

[2] Zinn-Justin J (1989)
{\em Quantum Field Theory and Critical Phenomena}.
Clarendon Press, Oxford, UK (fourth ed. 2002)

[3] Aref'eva I, Kulish P, Nissimov E, and Pacheva S (1978)
{\em Infinite Set of Conservation Laws of the Quantum Chiral Field in
Two-Dimensional Space-Time}.
Steklov Math. Inst. report LOMI E-1-1978, Nauka, St. Petersburg, Russia

[4] Aref'eva I, Krivoshchekov V, and Medvedev P (1980)
{\em 1/N Perturbation Theory And Quantum Conservation Laws For Supersymmetric 
Chiral Field}.
Theoretical and Mathematical Physics 40:3; 42:306 

[5] Aref'eva I, Nissimov E, and Pacheva S (1980)
{\em BPHZL Renormalization of $1/N$ Expansion and Critical Behaviour of the
Three-Dimensional Chiral Field}. Communications in Mathematical Physics 71:213

[6] Nissimov E, and Pacheva S (1981)
{\em Renormalization of the $1/N$ Expansion and Critical Behavior of
$(2+1)$-Dimensional Supersymmetric Non-Linear Sigma-Modles''},
Letters in Mathematical Physics 5:333-340; {\em Phase Transitions and 
$1/N$ Expansion in $(2+1)$-Dimensional Supersymmetric Sigma-Models},
Letters in Mathematical Physics 5:67-74

[7] Nissimov E, and Pacheva S (1984) {\em Dynamical Breakdown and Restoration of
Parity Versus Axial Anomaly in Three Dimensions}. Physics Letters B146:227-232;
\textsl{ibid} B160:431

[8] Klebanov I, and Polyakov A (2002), 
{\em AdS Dual of the Critical O(N) Vector Model}.
Physics Letters B550:213-219

\newpage
\begin{center}
{\Large \textbf{BPHZL Renormalization}}
\end{center}

The mass-independent (``soft-mass'') renormalization scheme of
of Zimmermann-Lowenstein [1,2] is based on the 
standard \underline{BPHZ} (Bogoliubov-Parasiuk-Hepp-Zimmermann) momentum-space 
subtraction procedure (the former is called 
\underline{{\em BPHZL renormalization}} scheme for short).
The general idea is to perform all subtractions in the
integrands of ultraviolet (UV) divergent \underline{Feynman diagrams} at 
zero external momenta and at {\em zero values of the mass parameters} except 
for those which by naive power counting would give rise to infra-red (IR) 
divergences, so that the latter subtractions are performed at zero external 
momenta but at {\em non-zero values of the mass parameters}. 

Technically, this is accomplished in the following way:

(a) One rescales temporarily all dimensionfull (mass) parameters $M$ entering 
the propagators and vertices of a diagram  $M \to s^{d_M}M$ where $d_M$ is
the canonical mass dimension of $M$ and at the end of the subtraction procedure
the auxiliary parameter $s$ is set to $s=1$.

(b) For the masses in the propagators of the fundamental bosonic ($\vp$) and
fermionic ($\psi$) matter fields one assigns temporarily a slightly more complex 
dependence on the auxiliary parameter $s$:
\br
-i\Bigl\lb (sm_{\vp} + (1-s)\m)^2 + P^2 (p,k)\Bigr\rb^{-1} \;\; ,\;\;
-i\( sm_{\psi} - \g^{\n} P_{\n} (p,k)\)
\Bigl\lb (sm_{\psi} + (1-s)\m)^2 + P^2(p,k)\Bigr\rb^{-1}
\nonu
\er
where $P(p,k)$ is a linear combination of external $\{ p\}$ and internal
$\{ k\}$ momenta,
$\m$ is arbitrary renormalization mass scale and again at the end of
the subtraction procedure one sets $s=1$.

(c) The momentum space Taylor subtraction operators $\t^{\d(\G),\r(\G)}$
in the fundamental \underline{``forest formula''} of the recurrsive 
\underline{BPHZ} subtraction scheme,
acting on the integrand of a UV-divergent (sub)diagram $\G$, are now defined as: 
$1 - \t^{\d(\G),\r(\G)} = 
\( 1 - t^{\r(\G)-1}_{\{ p\},s-1}\) \( 1 - t^{\d(\G)}_{\{ p\},s}\)$.
Here $\d(\G)$ and $\r(\G)$ are the UV and IR indices of the (sub)diagram $\G$
determined from the asymptotic behaviour of its integrand for large
internal momenta, and for small internal momenta at vanishing external momenta
and all masses set to zero, respectively.
$t^n_{x,y}$ denotes the usual Taylor subtraction operator:
$t^n_{x,y} F(x,y) \equiv \sum_{k,l=0\, ,\, k+l \leq n}^n
\frac{x^k}{k!} \frac{y^l}{l!} \frac{\pa^{k+l}F}{\pa x^k \pa y^l}\bv_{x=0,y=0}$.

\underline{BPHZL renormalization} has found a non-trivial application in the
systematic renormalization of non-perturbative \underline{large-$N$ expansions}
of \underline{nonlinear sigma-models} and their \underline{supersymmetric}
extensions [3,4] which look ``non-renormalizable'' from the point of view of
naive \underline{perturbation theory} w.r.t. coupling constants and which 
display rich phase structure with various ``low-temperature'' phases containing 
massless particles.

[1] Lowenstein J, and Zimmerman W (1975)
{\em On the Formulation of Theories with Zero-Mass Propagators}.
Nuclear Physics B86:77-103

[2] Lowenstein J (1976)
{\em Convergence Theorems for Renormalized Feynman Integrals with Zero-Mass
Propagators}.
Communications in Mathematical Physics 47:53-68

[3] Aref'eva I, Nissimov E, and Pacheva S (1980)
{\em BPHZL Renormalization of $1/N$ Expansion and Critical Behaviour of the
Three-Dimensional Chiral Field}. Communications in Mathematical Physics 71:213

[4] Nissimov E, and Pacheva S (1981)
{\em Renormalization of the $1/N$ Expansion and Critical Behavior of
$(2+1)$-Dimensional Supersymmetric Non-Linear Sigma-Modles''},
Letters in Mathematical Physics 5:333-340

\end{document}